# High temperature dust condensation around an AGB star: evidence from a highly pristine presolar corundum

Aki Takigawa[1,2], Rhonda M. Stroud[3], Larry R. Nittler[4], Conel M. O'D Alexander[4], and Akira Miyake[2]

[1]The Hakubi Center for Advanced Research, Kyoto University, Kitashirakawa-Oiwakecho, Sakyo, Kyoto 606-8502, Japan

[2]Division of Earth and Planetary Sciences, Kyoto University Kitashirakawa-Oiwakecho, Sakyo, Kyoto 606-8502, Japan

[3]Naval Research Laboratory, Code 6360, Washington, DC 20375, USA

[4]Department of Terrestrial Magnetism, Carnegie Institution of Washington, Washington, DC 20015, USA

**Abstract**

Corundum (α–$Al_2O_3$) and amorphous or metastable $Al_2O_3$ are common components of circumstellar dust observed around O-rich asymptotic giant branch (AGB) stars and found in primitive meteorites. We report a detailed isotopic and microstructural investigation of a unique presolar corundum grain, QUE060, identified in an acid residue of the Queen Alexandra Range 97008 (LL3.05) meteorite. Based on its O and Mg isotopic compositions, this 1.4 μm diameter grain formed in a low or intermediate mass AGB star. It has four developed rhombohedral {011} faces of corundum and a rough, rounded face with cavities. High Mg contents (Mg/Al>0.004) are due to the decay of radioactive $^{26}Al$. No spinel ($MgAl_2O_4$) inclusions that might have exsolved from the corundum are observed, but there are several high-Mg domains with modulated structures. The subhedral shape of grain QUE060 is the first clear evidence that corundum condenses and grows to micrometer sizes in the extended atmospheres around AGB stars. The flat faces indicate that grain QUE060 experienced little modification by gas-grain and grain-grain collisions in the interstellar medium (ISM) and solar nebula. The Mg distribution in its structure indicates that grain QUE060 has not experienced any severe heating events since the exhaustion of $^{26}Al$. However, it underwent at least one very transient heating event to form the high-Mg domains. A possible mechanism for producing this transient event, as well as the rough surface and cavity, is a grain-grain collision in the ISM. These results indicate that grain QUE060 is the most pristine circumstellar corundum studied to date.

**Introduction**

Corundum, the only thermodynamically stable phase of alumina ($Al_2O_3$), is one of the most refractory dust species expected to condense from a gas of the solar composition (e.g., Ebel 2006). Mid-infrared (MIR) spectroscopic observations have revealed that more than 90% of semi-regular variables and 20% of Mira variables, which are on the asymptotic giant branch (AGB), show single peaks at 13 μm (e.g., Speck et al. 2000; Sloan et al. 2003). Corundum is the most plausible dust species capable of producing this 13 μm feature (Zeidler et al. 2013; Takigawa et al. 2015). Amorphous or metastable alumina, which shows a broad MIR spectral feature peaking at 11-12 μm, is roughly as abundant as silicate dust around many O-rich AGB stars (e.g., Little-Marenin and Price 1986; Speck et al. 2000; Sloan et al. 2003). Recent radio observation of such an alumina-rich star, W Hya, shows that AlO gas molecules efficiently condense to dust while most SiO molecules remain in gas (Takigawa et al. 2017). These authors also suggested that alumina dust grows and accumulates near W Hya from the observed gas distribution and dust mass. However, dust formation in the extended atmospheres of alumina-rich AGB stars is poorly understood because of the limited mineralogical and crystallographic information that is available. It is not even clear if circumstellar corundum is a direct condensate or if it forms by post-condensation transformation of amorphous or metastable alumina that formed in very rapid cooling gases.

Presolar dust grains that must have formed prior to the birth of the Solar System are found in primitive chondritic meteorites and cometary dust, and have isotopic compositions indicating origins around evolved stars, such as AGB stars and supernovae. They may not be fully representative of all circumstellar dust grains, but because they can be studied in the laboratory with state-of-the-art microanalytical instrumentation, they can provide detailed quantitative information about the histories of individual dust grains that cannot be obtained by astronomical observations.

Presolar $Al_2O_3$ grains with isotopic compositions indicating AGB star origins have been identified in primitive chondrites (e.g., Hutcheon et al 1994; Nittler et al. 1997; Takigawa et al. 2014). The morphologies and crystal structures of presolar grains from AGB stars will reflect the condensation conditions in the circumstellar envelopes of their parent stars (e.g., pressure, temperature, cooling rate and chemical composition of the envelope), but may also record alteration in the interstellar medium (ISM) and protosolar disk. Refractory presolar grains like SiC, graphite, $Al_2O_3$ and $MgAl_2O_4$ (spinel) can be isolated from their host meteorites via physical and chemical processes (Amari et al. 1994; Bernatowicz et al. 2003; Cody et al. 2002). However, most of the $Al_2O_3$ grains in meteorite residues have solar origins (Nittler et al. 1997, 2008; Choi et al. 1998; Takigawa et al. 2014), and presolar $Al_2O_3$ grains thus must be identified by secondary ion mass spectroscopic (SIMS) isotopic measurements. Since SIMS sputters and alters grain surfaces, determining grain morphologies requires detailed scanning electron microscope (SEM) examination of each grain before the SIMS measurements. Morphological studies of pristine presolar $Al_2O_3$ grains are thus severely limited in number (Choi et al. 1998; Takigawa et al. 2014). Transmission electron microscopy (TEM) studies of presolar oxide grains were made possible with the development of focused ion beam (FIB) methods (Stroud et al. 2004), but the number of such studies is also small (Stroud et al. 2004, 2007; Zega et al. 2011; Zega et al. 2014). The microstructures of presolar $Al_2O_3$ grains reported to date include an amorphous grain, several corundum grains, and a crystalline $Al_2O_3$ tentatively identified as having a hexagonal crystal structure (Stroud et al. 2004, 2007).

Here we report on a unique subhedral presolar $Al_2O_3$ grain found in a residue of the Queen Alexandra Range (QUE) 97008 ordinary chondrite. From a systematic study of $Al_2O_3$ grains in acid residues that combined SEM, SIMS, FIB, and TEM, in that order, we discuss the origins of the morphology and interior microstructures, and thermal history of the grain.

**Experimental**

A residue of the primitive (LL3.05) QUE 97008 meteorite was prepared using the CsF procedure described in Cody et al. (2002) and Cody and Alexander (2005) to remove the bulk of the meteoritic material, follow by treatment with boiling perchloric acid, to destroy carbonaceous material and chromite ($FeCr_2O_4$). Droplets of the residue suspended in isopropanol and MilliQ water were deposited onto a high purity Au substrate so that on drying the grains were sparsely dispersed over its surface. Seventy-two $Al_2O_3$ grains were initially identified by energy dispersive X-ray spectroscopy (EDS) in a field-emission (FE-) SEM (JEOL JSM-6500F) at the Carnegie Institution of Washington (CIW) and then imaged from five directions by tilting and rotating the SEM stage. Electron backscatter diffraction (EBSD) measurements of grain surfaces were also made with a FIB-SEM (FEI Nova 600) with an HKL EBSD system at the US Naval Research Laboratory (NRL) on all grains to determine their crystal structures. Details of the EBSD method are described in Takigawa et al. (2014). Additional $Al_2O_3$ grains in acid residues of the ordinary chondrites Semarkona (LL3.0, 17 grains), Bishunpur (LL3.15, 23 grains), and Roosevelt County 075 (H3.1, 37 grains), previously identified by Takigawa et al. (2014), were also subjected to isotopic analysis.

The isotopic measurements were performed with the Cameca NanoSIMS 50L ion microprobe at CIW. The O isotopes of 149 $Al_2O_3$ grains were measured with a ~100 nm $Cs^+$ primary ion beam rastered over each of the grains with simultaneous detection of the secondary ions of $^{12}C^-$, $^{16}O^-$, $^{17}O^-$, $^{18}O^-$, $^{24}Mg^{16}O^-$, and $^{27}Al^{16}O^-$. The primary beam sputtering of the grains was minimized by halting the isotopic measurements as soon as isotopic anomalies were clearly detected. This resulted in increased counting statistical errors for the presolar grains compared to the grains of Solar System origin (i.e., those without clear O isotopic anomalies). An ~400 nm $O^-$ primary ion beam was used to measure the Mg-Al isotopic compositions of the presolar and some solar $Al_2O_3$ grains. Secondary ions of $^{24}Mg^+$, $^{25}Mg^+$, $^{26}Mg^+$, and $^{27}Al^+$ were simultaneously measured. Micron-sized Burma spinel grains were used as a standard to correct for the instrumental mass fractionation and to determine the relative sensitivity factor of secondary Mg and Al ions. The standard $^{26}Mg/^{24}Mg$ of 0.13932 (Catanzaro et al. 1966) and a sensitivity factor of 1.16 were used to calculate the initial $^{26}Al/^{27}Al$ ratio.

Ultra-thin sections of a few identified presolar grains were prepared with the FIB-SEM at NRL. TEM studies were carried out with field-emission scanning transmission electron microscopes (STEM) at NRL (JEOL JEM-2200FS) and Kyoto University (JEOL JEM-2100F equipped with JED 2300D). Chemical compositions were measured by STEM-EDS. The K-factor method (Cliff and Lorimer 1975) was used for quantification of the EDS data. We focus here on the TEM data from one particularly interesting grain, QUE060; results for the other grains will be reported elsewhere.

**Results**

*Isotopic compositions*

The O isotopic measurements revealed six presolar grains from the acid residue of QUE 97008, and one from that of RC075 (Fig. 1 and Table 1). No new presolar grains were identified from the residues of Semarkona and Bishunpur. Combined with the grains reported by Takigawa et al. (2014), the presolar grain fraction in $Al_2O_3$ grains of residues from Semarkona, Bishunpur, RC 075, and QUE 97008 are 0/25, 2/72, 8/87, and 6/72, respectively.

Grain QUE060 is enriched in $^{17}O$ compared to terrestrial by a factor of ~2.8, but depleted in $^{18}O$ by a factor of 25 (Table 1), which identifies it as a Group 2 grain according to the classification scheme for presolar oxide grains (Nittler et al. 1997). A large excess of $^{26}Mg$ was detected due to the decay of radioactive $^{26}Al$, with an inferred initial $^{26}Al/^{27}Al$ ratio of 0.0126(23) (Table 1). The O and Al-Mg isotopic compositions indicate that grain QUE060 either originated either in a ~5 $M_{Sun}$ AGB star undergoing hot-bottom processing (Lugaro et al. 2017) or a lower-mass (< 2 $M_{Sun}$) AGB star undergoing cool-bottom-processing (Nittler et al. 2008)

*Morphology and surface structure*

Figure 2 shows secondary electron images of grain QUE060 on an Au substrate that were acquired prior to modification of its surface morphology by the SIMS isotopic measurements. The mean of the longest and shortest dimensions of grain QUE060 is about 1.4 μm. Four flat faces are clearly observed on the grain (faces *a-d* in Fig. 2). Such flat and smooth faces were not observed for any other presolar $Al_2O_3$ grains examined here, nor have they been reported in previous studies (Choi et al. 1998; Takigawa et al. 2014). The area *e* shows a rough and rounded surface with no clear edge and seems to have cavities (Figs. 2B-D). The rough surface structure of area *e* is similar to the characteristic surface structures observed on most previously studied presolar $Al_2O_3$ grains (Choi et al. 1998; Takigawa et al. 2014).

The EBSD patterns taken prior to the SIMS measurements on multiple locations on face *b* (Fig. 2) were indexed to that of corundum. The EBSD analysis shows that at least the very surface layer (<30 nm in depth) sampled by the measurements, which was subsequently lost during the SIMS measurements, originally had the corundum crystal structure, which is consistent with that of grain interior as shown in the following section.

*Interior microstructures*

A FIB lift-out section of grain QUE060, cut perpendicular to the Au substrate along the dashed lines in Fig. 2F, was prepared for TEM analysis (Fig. 3). The SEM image in Fig. 2D was taken almost parallel to the TEM images in Figure 3. The flat face indicated by the arrow in Fig 3A corresponds to face *a* in Fig. 2. Comparing the initial shape and size (Fig. 2D) with those in the FIB-section (Fig. 3A), most of the grain volume was preserved after the SIMS measurements. A cavity seen on area *e* of Fig. 2 was also observed on the left part of the grain facing the Au substrate (Fig. 3A). The ion beam damage caused by the SIMS measurements and FIB lift-out cannot be responsible for the cavity formation because the cavity was not exposed to the electron and ion beams.

Very similar electron diffraction patterns of corundum (Fig. 3B) were obtained from most locations across the whole grain (the few exceptions will be described later), which indicates that this grain is a single crystal of corundum. From the EBSD pattern and the electron diffraction patterns along the three different zone axes, the faceted faces of *a*, *b*, *c*, and *d* in Fig. 2 are indexed to rhombohedral (011), (110), (101), and ($\bar{1}0\bar{1}$) faces, which are crystallographically identical {011} planes and generally called r-planes (Hartman 1980). Note that we took the smallest rhombohedral unit cell to describe the crystal faces and directions of corundum in this paper.

The bright-field (BF) TEM image in Fig. 3A can be seen to be darkening from right to left. Electron diffraction patterns taken at the center of the grain and the edge of the cavity showed a rotational difference of ~0.3 degrees. The gradual change of the contrast in the BF-TEM image from left to right (large scale contrast) is thus due to rotational differences of the crystallographic orientation. Both the BF and the dark-field (DF) TEM images (Fig. 3A and D) show small scale (<30 nm) brightness contrast throughout the grain, which indicates that distortions are distributed everywhere in the grain. The selected-area electron diffraction (SAED) patterns from the area b in Fig. 3A showed satellite spots (Fig. 3C), indicating modulated structures in the corundum crystal. These satellite spots were observed from several domains corresponding to the bright areas in the DF-TEM image (Fig. 3E). These domains are <130 nm in diameter.

The average Mg/Al ratio in the grain measured with STEM-EDS is 0.014(1) The detected Mg is essentially pure radiogenic $^{26}$Mg based on the measured Mg isotopic composition (Table 1) and the ratio obtained with EDS is consistent with the $^{26}$Mg/$^{27}$Al of 0.0126(15) determined by NanoSIMS. The Mg/Al ratios are not uniform within the grain, but are always greater than Mg/Al ~0.004(1) (2000 ppm Mg by weight), which is significantly higher than the Mg detection limit of 0.001 in our measurements. The domains showing satellite spots contain a higher amount of Mg (Fig. 3F). The Mg/Al ratios in such domains are up to 0.038(3), which is significantly higher than those of the surrounding regions. The DF image of satellite spots shows that the modulated structure is not present in the lower Mg-content regions (Fig. 3E).

**Discussion**

*Formation of grain QUE060*

The existence of presolar corundum grains does not rule out the possibility that the grains originally condensed as amorphous or metastable alumina and later transformed to corundum at >1000°C (Levin and Brandon 1998). Had grain QUE060 initially condensed as amorphous alumina around its parent AGB star, the faceted surfaces would have to have formed later by surface diffusion of molecules following crystallization. Because amorphous $Al_2O_3$ formation needs very rapid cooling of the gas (e.g., Dragoo and Diamond 1967), reheating is required for crystallization and surface diffusion. Heating events in the outflows of AGB stars beyond the dust formation regions are mainly due to shock wave propagation. These are transient events, whereas sustained high temperatures are needed for faceted face formation by surface diffusion. Melting or surface diffusion in the protosolar disk can be ruled out because isotopic exchange with the surrounding gas is very efficient at high temperatures and would have erased the isotopic anomalies. Thermal alteration in the parent body of QUE 97008 is also unlikely because QUE 97008 (LL3.05) shows very little evidence of aqueous alteration or thermal metamorphism (Grossman and Brearley, 2005). Hence, the faceted faces on grain QUE060 suggest that this grain originally condensed as crystalline corundum.

Corundum grains found naturally in the Earth or synthesized with flux methods in laboratories generally show well developed {111} faces (c-planes). However, the most developed faces theoretically predicted using attachment energies and from *ab-initio* calculations are {110} (Hartman 1980, 1989; Rhol and Gay 1995) as observed for grain QUE060. Hartman (1989) argued that the development of {111} faces is due to adsorption of atoms on the {111} faces, where the density of unsaturated bonds is highest, and that {110} faces should develop at high temperature and low supersaturation conditions. It is, therefore, most likely that the faceted faces observed on grain QUE060 developed during condensation and growth of crystalline $Al_2O_3$ around its parent AGB star. There is no evidence of reaction between grain QUE060 and the surrounding gas, which indicates that the grain remained and grew within the high temperature region near the central star where no other mineral was thermodynamically stable. The grain was then transported rapidly to the low temperature and gas density regions along with the accelerated outflow so that other refractory grains could not grow on it or form by gas-grain reaction.

### *Formation of the rough surface and cavities*

The uniqueness of the very flat {110} faces on grain QUE060 indicates that this grain did not spend a very long time in the diffuse ISM. The rough surface area on grain QUE060 (area *e* in Fig. 2) may be a secondary feature possibly formed by dust-gas or dust-dust collisions in the ISM or protosolar disk. Corundum is not dissolved even by more harsh acid treatments using HF, but amorphous and metastable alumina phases do dissolve into HF-HCl or $HClO_4$ (Takigawa et al. 2014). If partial amorphization or deformation of the crystal structure occurred as the result of such a collision, the damaged areas could have been dissolved during the isolation procedures.

A grain-grain collision (Jones et al. 1996) is also a possible mechanism for forming the cavity observed in the FIB section. In addition to forming a crater, a high velocity collision would have generated a shock wave that propagated through the grain (Jones et al. 1996). The rotational misorientation from left to right in the FIB section may be explained by such a shock wave propagating from the cavity.

### *Thermal history after $^{26}Al$ decay*

The inferred initial $^{26}Al/^{27}Al$ of grain QUE060 is about 0.01. Thus, the radiogenic $^{26}Mg$ produced by $^{26}Al$ decay should have initially occupied roughly one percent of the octahedral Al-sites in the corundum structure. The solubility limit of MgO in corundum is 130 ppm at 1600 °C (Miller et al. 2006) and the abundance of Mg in grain QUE060 (≥2000 ppm) is much higher than this. Since spinel ($MgAl_2O_4$) and corundum show complete immiscibility, if it had been able to achieve thermodynamic equilibrium after the decay of the $^{26}Al$ grain QUE060 would have decomposed to $^{26}Mg^{27}Al_2O_4$ and corundum by the following reaction:

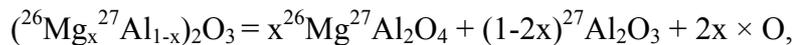
$$(^{26}Mg_x{}^{27}Al_{1-x})_2O_3 = x^{26}Mg^{27}Al_2O_4 + (1-2x)^{27}Al_2O_3 + 2x \times O,$$

where x denotes the initial ratio of $^{26}Al/^{27}Al$ when the grain condensed. If all $^{26}Mg$ diffused to form stoichiometric spinel, about 1.6 vol.% of the grain should have become spinel. No precipitate of Mg-bearing phases was observed even in the Mg-rich domains

- the crystal structures in the Mg-rich domains are corundum with the modulated structure. These observations indicate that there was at least one heating event to form the modulated structures by Mg diffusion, but the grain did not gain enough energy to exsolve spinel.

Based on the sizes, Mg/Al ratios, and distribution of the Mg-rich domains, $^{26}$Mg atoms could not have diffused in grain QUE060 by more than 0.1 μm. We estimated the maximum temperature that grain QUE060 could have experienced after the decay of $^{26}$Al using the activation energy and pre-exponential terms for Mg self-diffusion of 874 kJ and 74.6 cm$^2$/s, respectively (Sheng et al. 1992). Here we assume that the diffusion coefficients of Mg in corundum and spinel are similar (MacPherson et al. 1995). Magnesium diffuses 0.1 μm within 20 Myr at >500°C, 1 hour at >1100 °C, and 1 second at >1530 °C. These estimated temperatures and durations are maximum values because the distortions in the corundum structure due to Mg atoms formed by $^{26}$Al decay in octahedral sites would have enhanced the diffusion rate. It is unlikely that the grain had remained at a few hundred K for such a long time while retaining its O isotopic anomaly, and thus these results indicate that after decay of $^{26}$Al, the grain may have experienced a transient heating event. This event must have occurred in the ISM or the early solar system because the time scale for grain injection into the ISM after formation around an AGB star is much shorter than the 720,000 yr half-life of $^{26}$Al. A collisional event in the ISM that could also have been responsible for the crater formation discussed above is an attractive explanation for the heat source that enabled the Mg-diffusion.

*Implication for astromineralogy*

The faceted faces of the presolar corundum grain QUE060, indexed to {011} faces, provide strong evidence for the direct condensation of corundum around an AGB star. Because of its very refractory nature, corundum condenses from a hot gas within a few stellar radii from the central star, where acceleration of the stellar wind does not occur. The micron size of grain QUE060 indicates that the grain had to survive several pulsation periods to grow (Gobrecht et al. 2016). Small alumina grains (<0.1 μm) are transparent to the stellar radiation, but micron-sized grains efficiently scatter the stellar light and contribute more to the acceleration of the stellar wind (Höfner 2008). Hence, grain QUE060 formed and grew for at least several years in the extended atmosphere of its parent star before being accelerated away from it by stellar radiation. The observation of AlO gas in the atmosphere of an alumina-rich AGB star, W Hya, suggests that growth and accumulation of alumina dust contributes to the triggering of the acceleration that generates a stellar wind (Takigawa et al. 2017). Micron-sized presolar corundum grains like grain QUE060 thus may have been such triggers of dust-driven winds from alumina-rich AGB stars.

Acknowledgements: We thank Gary R. Huss for providing the samples of Semarkona, RC 075, and Bishunpur. This work was supported by JSPS KAKENHI Grant Number JP12J02495 (AT), and NASA Grants NNH16AC42I (RMS and CA) and NNX10AI63G (LRN and CA).

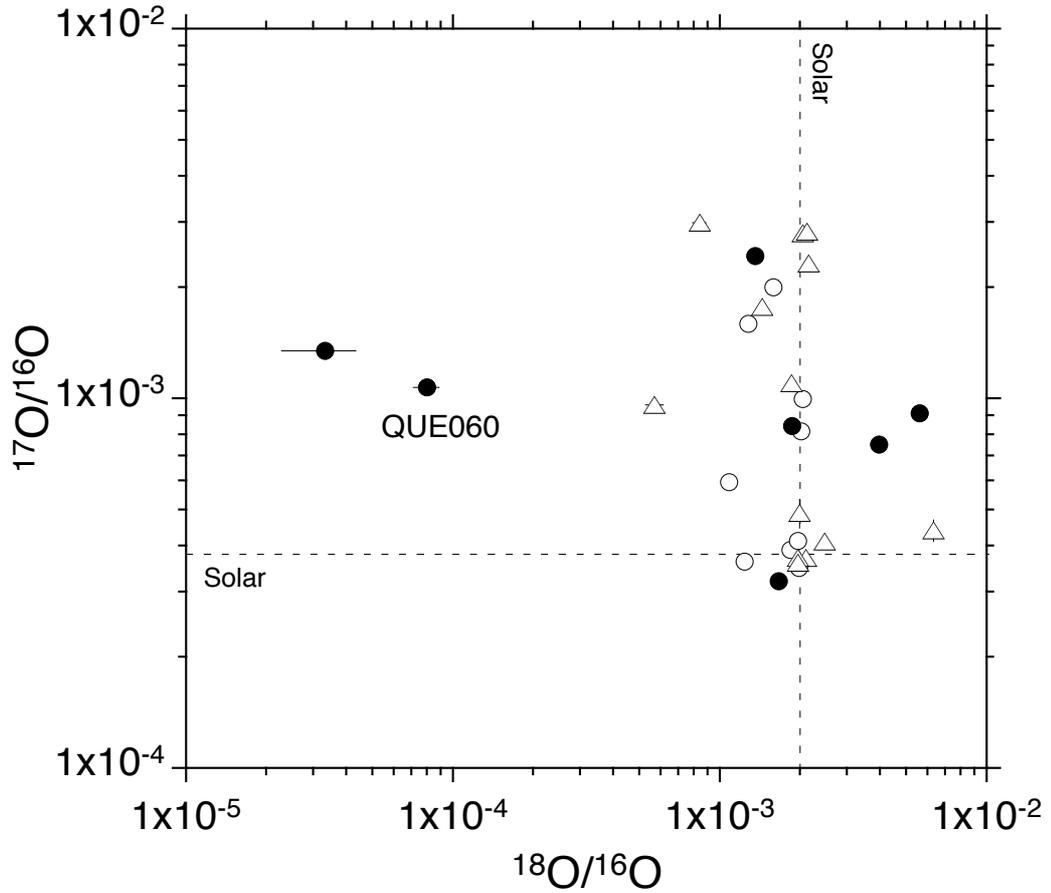

**Figure 1.** Oxygen isotopic compositions of presolar $Al_2O_3$ grains from ordinary chondrites RC 075, Bishunpur, Semarkona and QUE 97008. Filled circles show newly found presolar grains in this study and open circles and triangles are previously measured grains from RC 075, Bishunpur (Takigawa et al., 2014) and those from Semarkona and Bishunpur (Choi et al. 1998), respectively.

Table 1: Isotopic compositions of the identified presolar grains[a].

| | Meteorite | $^{17}O/^{16}O$ | $^{18}O/^{16}O$ | $^{25}Mg/^{24}Mg$ | $^{26}Mg/^{24}Mg$ | $^{24}Mg/^{27}Al$ | $^{26}Al/^{27}Al_0$ |
|---|---|---|---|---|---|---|---|
| QUE018 | QUE 97008 | 8.41 (6) × 10⁻⁴ | 1.87 (2) × 10⁻³ | 0.123 (1) | 0.140 (1) | 2.88 (1) × 10⁻² | 1.80 (1) × 10⁻⁵ |
| QUE053 | QUE 97008 | 2.43 (4) × 10⁻³ | 1.36 (4) × 10⁻³ | 0.213 (48) | 0.200 (5) | 3.33 (33) × 10⁻⁴ | 2.03 (52) × 10⁻⁵ |
| QUE060 | QUE 97008 | 1.07 (1) × 10⁻³ | 7.99 (1.69) × 10⁻⁵ | 0.113 (3) | 99.0 (8.6) | 1.27 (11) × 10⁻⁴ | 1.26 (23) × 10⁻² |
| QUE067 | QUE 97008 | 9.13 (15) × 10⁻⁴ | 5.60 (4) × 10⁻³ | 0.129 (15) | 6.73 (31) | 4.63 (20) × 10⁻⁴ | 3.07 (19) × 10⁻³ |
| QUE088 | QUE 97008 | 4.14 (7) × 10⁻⁴ | 2.16 (1) × 10⁻³ | 0.135 (14) | 0.110 (14) | 1.80 (74) × 10⁻⁴ | - |
| QUE124 | QUE 97008 | 7.52 (12) × 10⁻⁴ | 3.96 (2) × 10⁻³ | 0.176 (86) | 263 (53) | 2.95 (59) × 10⁻⁵ | 7.75 (2.19) × 10⁻³ |
| QUE137 | QUE 97008 | 1.35 (1) × 10⁻³ | 3.31 (1.36) × 10⁻⁵ | 0.139 (7) | 7.40 (15) | 2.19 (4) × 10⁻³ | 1.59 (4) × 10⁻² |
| RC075 58-02 | RC 075 | 3.21 (6) × 10⁻⁴ | 1.66 (1) × 10⁻³ | 0.138 (31) | 0.0810 (245) | 5.67 (50) × 10⁻⁵ | - |

[a] all errors are 1 σ.

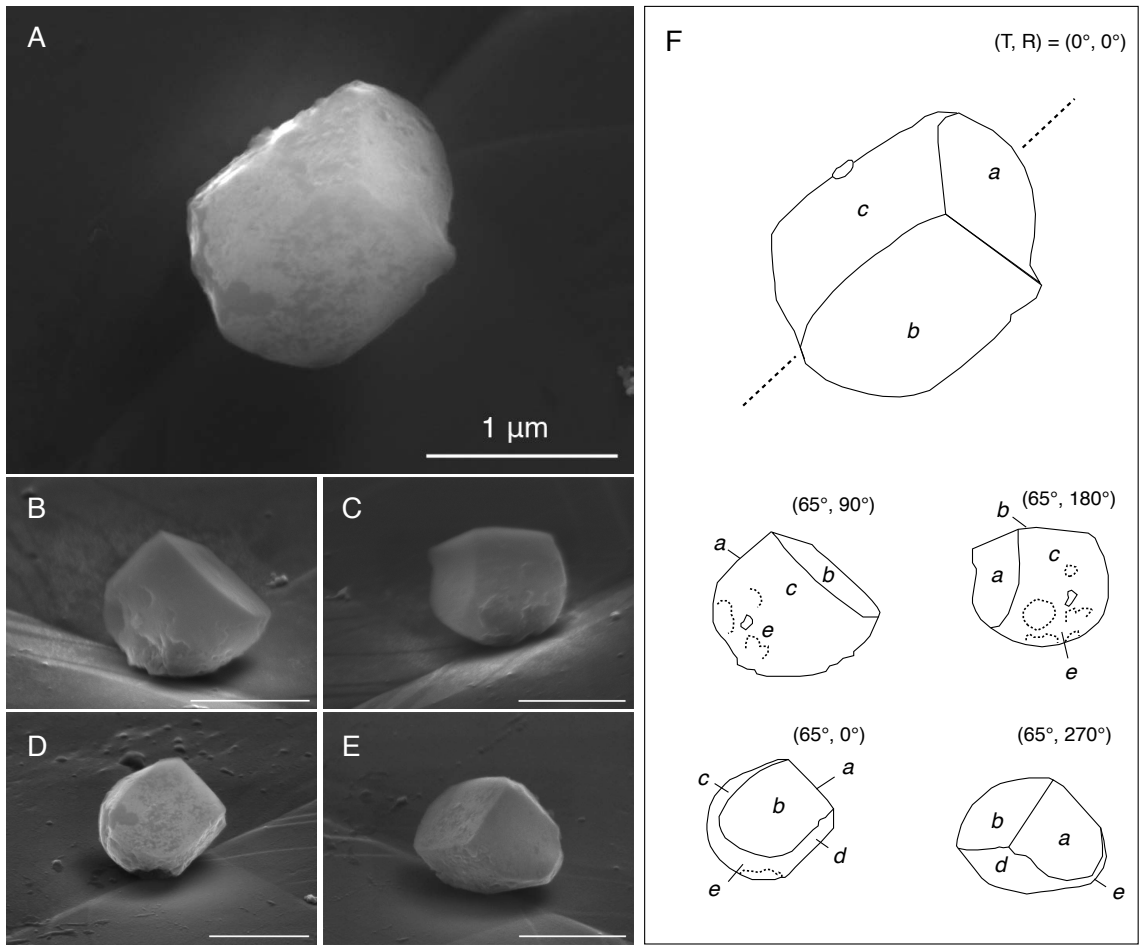

**Figure 2**: Secondary electron images (A-E) and outlines (F) of grain QUE060 taken from five different directions tilting (T = 0 and 65°) and rotating (R = 0, 90, 180, and 270°) the SEM stage. The tilting and rotating angles of Figs. A-E are shown in Fig. F. A FIB section was prepared along the dash lines in (T, R) = (0°, 0°) image of Fig. F. A small subgrain (<100 nm) is attached to the area *e*. The faces of *a*, *b*, *c*, and *d* were indexed to rhombohedral $(011)$, $(110)$, $(101)$, and $(\bar{1}0\bar{1})$ faces of corundum.

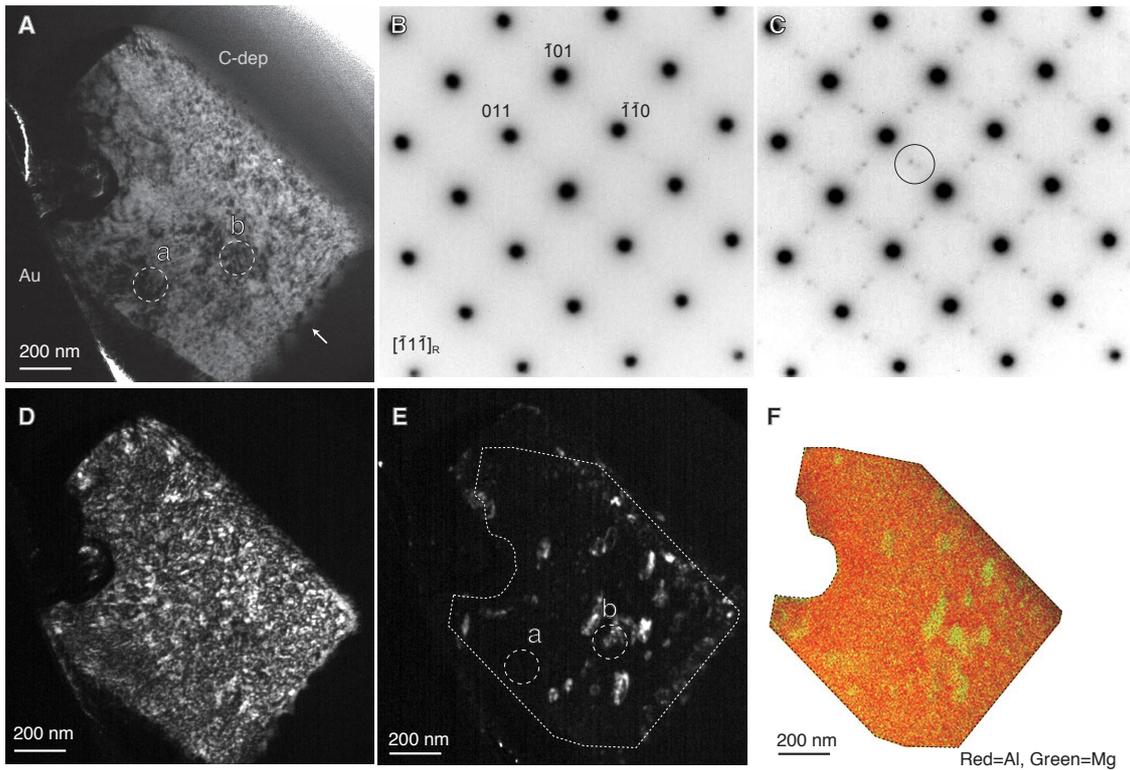

**Figure 3**: (A) BF-TEM image of the FIB section of grain QUE060. An arrow indicates flat face a in Fig. 2. SAED patterns of the areas indicated by the dashed circles a and b are shown in panels B and C, respectively. They are taken in the $[\bar{1}1\bar{1}]$ orientation of corundum (written in the rhombohedral setting). (D) DF-TEM image using a diffraction 011 spot. (E) DF-TEM image from spots indicated by a circle in panel C. (F) Al (red) and Mg (green) overlay map obtained with STEM- EDS.